# Consistency of Social Sensing Signatures Across Major US Cities

Aiman Soliman[1,2] Kiumars Soltani[1,4] Anand Padmanabhan[1,2,3,4] Shaowen Wang[1,2,3,4]

[1] CyberGIS Center for Advanced Digital and Spatial Studies
[2] National Center for Supercomputing Applications (NCSA)
[3] Department of Geography and Geographic Information Science
[4] Illinois Informatics Institute
University of Illinois at Urbana-Champaign, Champaign, IL, USA

**Previous studies have shown that Twitter users have biases to tweet from certain locations (locational bias) and during certain hours (temporal bias). We used three years of geo-located Twitter Data to quantify these biases and test our central hypothesis that Twitter users' biases are consistent across US cities. Our results suggest that temporal and locational bias of Twitter users are inconsistent between three US metropolitan cities. We derive conclusions about the role of the complexity of the underlying data producing process on its consistency and argue for the potential research avenue for Geospatial Data Science to test and quantify these inconsistencies in the class of organically evolved Big Data.**

**Introduction**

Recently, we have witnessed an increase in the volume of digital footprints around urban environments. Such information has become increasingly crucial to understand the fast evolving urban landscapes and augments traditional high latency on-site survey methods. However, a major problem that is common among the organically evolved big data sets is the lack of information about its consistency [1]. Unlike traditional measurements, where data collection protocols guarantee its consistency and reproducibility, the data generation processes underlying most of big data are usually unknown. This lack of knowledge about the consistency of ambient geospatial big data has resulted in limitations, which over generalize the results beyond the presented case studies. We highlight this problem by testing the hypothesis of consistency of two metrics of Twitter users; namely the user preference to engage ('tweet') around certain landuse types, and the Twitter users' circadian rhythms across major metropolitan US cities.

Previous research demonstrated the possibility of inferring urban landuse from Twitter data based on the analysis of individual user mobility patterns [2]. In this regard, the underlying process generating the spatial temporal patterns of Twitter data is a convolution of two processes. First the process of engagement with the technology (tweeting), and second, the mobility patterns of technology users. Although human mobility is highly predictable and consists of a few key locations (e.g., home, work, etc.) [3], Twitter user biases are not well understood [4]. In this research work, we compared Twitter user biases toward tweeting from certain landuse type/urban activity and during certain times of the day. Our main guiding hypothesis is that there are no significant differences of the Twitter user biases across US cities. We tested this hypothesis using detailed landuse maps in different US cities and used it to quantify the consistency of a three year collection of geolocated Twitter data.

Here we demonstrate that assumptions made of about the big geolocated data sets must be proven true before incorporating it into larger studies. Specifically, we demonstrate that the process of geolocated tweet production is not spatially consistent, even when mined from the same source (e.g. the Twitter API).

**Datasets and Methods**

      Geotagged tweets were obtained using Twitter's streaming API from January 2013 to January 2016 (~ 2.42 billions tweets for contiguous United States). From these we selected tweets within the geographic bounding boxes of Chicago (39 million tweets), Manhattan (18 million tweets), and San Diego (8 million tweets). We filtered the Twitter data to remove duplicate records. In addition, we removed the tweets of Twitter users who made less than 10 tweets per year, were active for less than 30 days, or exceeded the 99% percentile for speed between consecutive tweets.

      We used parcel level landuse maps recently released from the New York City Department of City Planning, San Diego Land Layers (SANDAG's), and landuse inventory for Northeastern Illinois to retrieve landuse types associated with each tweet collected in the cities of New York, San Diego and Chicago respectively. We assigned each tweet to the nearest landuse parcel using a scalable point in nearest polygon algorithm. Landuse types were grouped into twelve activity classes using a legend, which is popular in social studies [5]. We applied a DBSCAN clustering function using a search window of 0.00225 degrees and a minimum of three points to identify 884,737 key locations from 163,340 users in the city of Chicago. Similarly we identified 192,934 key locations from 47,356 users in in San Diego and 503,223 locations from 132,546 users in the island of Manhattan. The spatial clustering was done to identify significant key locations as indicated by multiple tweets at the same location and avoid random tweets. We labeled each key location using the dominant landuse associated with its tweets.

      Twitter users are known to exhibit biases, sending tweets around certain locations (e.g., tweeting at home) or during certain hours of the day [4]. In the absence of locational bias, the distribution of different landuse types in Twitter data should resembles their abundance in the city. However, if a preferential bias exists, some landuse types will be more common in Twitter data compared to their relative weight (share) in the city. Our first metric assess the locational bias by measuring the ratio of landuse abundance in Twitter data compared to the city landuse map. The first weight measure is the ratio of the number of Twitter clusters (users' key locations) labeled with a certain landuse type to the total number of clusters, while the latter measure is a relative ratio of occupied surface area of a certain landuse type to the total surface area of the city. We fitted a linear model to the relation between weights of different landuse types independently for each city. A slope of one indicates an equal abundance weight of landuse types in both Twitter and the landuse map (null hypothesis). In addition, we extracted the hourly volume of tweets associated with different landuse types in each of the studied cities. Our second metric quantified the similarity of these signals across cities by calculating the pairwise distance based on the Dynamic Time Warping algorithm. Our hypothesis that if Twitter data is consistent then activity signatures of the same landuse type should be closely related to each other across different cities.

**Results and Discussions**

Figure 1 (a) (b) summarizes the dominant landuse types of all the Twitter users' key locations grouped by location rank, where rank one shows the most frequently tweeted from location for all users in the cities of Chicago and San Diego respectively. It is clear from the figure that there is not a strong association between the rank and the dominant landuse type. One simple explanation is that the factors that determine the ranking of a user's key locations varies from individual to individual and can be regarded,for a large ensemble of users, as a random variable. Under this assumption, the only controlling factor of the importance of a landuse type for a Twitter user is its abundance in the city. In order to account for variable landuse composition between cities we plotted the relative weight of each landuse type within the Twitter data against its surface area on the map (figure 1 (c)). If the random hypothesis is true then we would expect the linear plots to align with the equal weight line (slope of 1).

The plot suggests that the location bias is limited in both New York and Chicago (linear slope close to one and $r^2 \cong 0.75$. However, Twitter users in San Diego showed a significant bias and much lower rates when tweeting from home. This disproportional bias results in outliers that alter the slope of the line and result in slope of approximately 0.5.

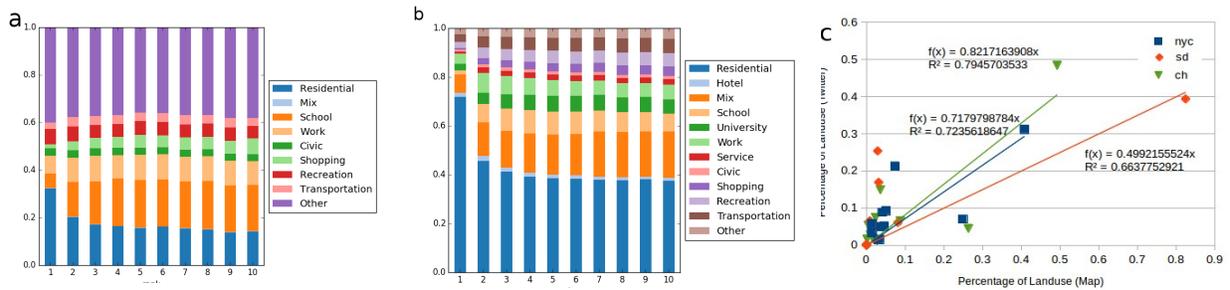

**Figure 1: Normalized landuse composition for top tweeted from locations grouped by rank (number of tweets) for all users in the city of (a) San Diego and (b) Chicago; legend colors are not the same between figures; ( c) the regression relation between landuse weights in Twitter and landuse maps; the no-locational bias of a Twitter user will correspond to equal weights in both datasets (slope of 1).**

The hourly change in the volume of tweets shows a distinct signature for each landuse type (Figure 2). For example, educational landuse types have a characteristic rise of activity during the early hours of the day followed by a strong decline of activity after 3 pm. We tested the hypothesis that social signatures of the same landuse type should be similar and consistent across cities. This hypothesis is justified by the universality of daily routines associated with different landuse types. However, the dendrogram (Figure 2c) based on the similarity of temporal signatures shows inconsistencies. Although the social signatures from San Diego and Chicago are very similar for most of landuse types, the social signals of different landuse types from Manhattan island are indistinguishable. We speculate that high degree of landuse mixing and high urban development intensity resulted in convoluted signals that combines multiple urban activity characteristics (results are not presented).

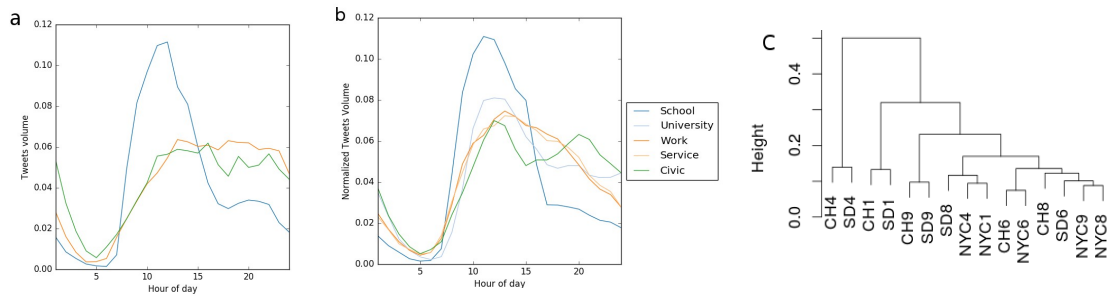

**Figure 2: Normalized volume of tweets per hour during weekday for the city of (a) San Diego and (b) Chicago; (c) hierarchical clustering of activity signatures based on a Dynamic Time Warping similarity distance where CH is chicago, SD is San Diego and NYC is New York City (Manhattan) and 1 is residential, 4 is educational/schools, 6 is work, 8 is civic and 9 is shopping.**

## Conclusions

We tested a general hypothesis about the consistency of Geospatial Big Data using the example of geolocated Twitter data collected over the course of three years at three major US cities. Two metrics were calculated to evaluate the users' preference to tweet from certain locations (locational bias) and during certain hours of the day (temporal bias). The results demonstrate that users from the Eastern US (Chicago and Manhattan) have less locational bias than users from San Diego who are less engaged with Twitter while at home. From another perspective, users from San Diego and Chicago have a consistent circadian rhythms (temporal bias) compared to users from Manhattan likely because the high degree of urban intensification in Manhattan island is not accounted for. These results shed light on the fact that data producing processes of geolocated Twitter data--and likely other organically evolved Geospatial Big Data-- are usually unknown and too complex to be approximated by a set of fixed assumptions. Therefore, it is a potential research avenue for data science to test hypotheses and expand our knowledge about the underlying production processes of different big data sets, which consequently will allow for more robust application of data mining techniques.

## Acknowledgements

This material is based in part upon work supported by the U.S. National Science Foundation under grant numbers: 1047916, 1429699, and 1443080. The authors would like to thank Austin Davis whose comments helped improve and clarify this manuscript.